\documentclass[10pt,conference]{IEEEtran}
\IEEEoverridecommandlockouts
\usepackage{cite}
\usepackage{amsmath,amssymb,amsfonts}
\usepackage{algorithm}
\usepackage{algpseudocode}
\usepackage{graphicx}
\usepackage{textcomp}
\usepackage{xcolor}
\usepackage{verbatim}
\usepackage{amsthm}
\def\BibTeX{{\rm B\kern-.05em{\sc i\kern-.025em b}\kern-.08em
    T\kern-.1667em\lower.7ex\hbox{E}\kern-.125emX}}
\usepackage{graphicx}
\usepackage{subcaption}    

\begin{document}

\title{Interplay between Sensing and Communication in Cell-Free Massive MIMO with URLLC Users
}
\author{\IEEEauthorblockN{
Zinat Behdad\IEEEauthorrefmark{1}, \"Ozlem Tu\u{g}fe Demir\IEEEauthorrefmark{2}, Ki Won Sung\IEEEauthorrefmark{1}, and Cicek Cavdar\IEEEauthorrefmark{1}
\thanks{Results incorporated in this paper received funding from the ECSEL Joint Undertaking (JU) under grant agreement No 876124. The JU receives support from the EU Horizon 2020 research and innovation programme and Vinnova in Sweden. }}
\IEEEauthorblockA{\IEEEauthorrefmark{1}Department of Computer Science, KTH Royal Institute of Technology, Stockholm, Sweden (\{zinatb, sungkw, cavdar\}@kth.se)}
\IEEEauthorblockA{\IEEEauthorrefmark{2}Department of Electrical-Electronics Engineering, TOBB ETU, Ankara, Türkiye  (ozlemtugfedemir@etu.edu.tr)}
}
\maketitle
\begin{abstract}
This paper studies integrated sensing and communication (ISAC) in the downlink of a cell-free massive multiple-input multiple-output (MIMO) system with multi-static sensing and ultra-reliable low-latency communication (URLLC) users. We propose a successive convex approximation-based power allocation algorithm that maximizes energy efficiency while satisfying the sensing and URLLC requirements. In addition, we provide a new definition for network availability, which accounts for both sensing and URLLC requirements. The impact of blocklength, sensing requirement, and required reliability as a function of decoding error probability on network availability and energy efficiency is investigated. The proposed power allocation algorithm is compared to a communication-centric approach where only the URLLC requirement is considered. 
It is shown that the URLLC-only approach is incapable of meeting sensing requirements, while the proposed ISAC algorithm fulfills both sensing and URLLC requirements, albeit with an associated increase in energy consumption. This increment can be 
reduced up to 75\% by utilizing additional symbols for sensing.
It is also demonstrated that 
larger blocklengths enhance network availability and offer greater robustness against stringent reliability requirements.
\end{abstract}

\section{Introduction}
Integrated sensing and communication (ISAC) aims to enhance spectral/hardware efficiency and is expected to enable various use cases for autonomous vehicles, smart factories, and other location/environment-aware scenarios in 6G communication networks \cite{liu2022integrated,zhou2022integrated}. Ultra-reliable low-latency communication (URLLC) is required in most of these use cases such as industrial Internet-of-Things (IIoT) and vehicle-to-everything (V2X) communications. In a V2X network, an ISAC system can detect targets of interest, such as pedestrians and vehicles, and deliver sensing information reliably with a minimal delay for URLLC users such as autonomous vehicles \cite{ding2022joint}. Therefore, studying both technologies together is vital to enable mission-critical applications in future wireless networks.

To the best of our knowledge, there are few works that jointly consider URLLC and ISAC.
In \cite{ding2022joint}, a joint precoding scheme is proposed to minimize transmit power, satisfying sensing and delay requirements. Moreover, joint ISAC beamforming and scheduling design is addressed in \cite{zhao2022jointbeam} for periodic and aperiodic traffic. 
The existing works consider a single ISAC base station, while cell-free massive MIMO (multiple-input multiple-output) with multiple distributed access points (APs) has been shown to facilitate URLLC \cite{lancho2023cell, ren2020joint, nasir2021cell} and ISAC\cite{sakhnini2022target,behdad2022power} 
implementations. This motivates us to study the interplay between URLLC and ISAC in cell-free massive MIMO systems. Specifically, we aim to address the power allocation problem so that both sensing and URLLC requirements are satisfied. For most URLLC applications, such as V2X and IIoT, short codewords are needed to satisfy latency constraints where codes with short blocklengths, e.g., 100 symbols are employed. Therefore, the finite blocklength regime needs to be considered to model decoding error probability (DEP) \cite{ozger-fblock2018, sun2018optimizing, lancho2023cell}. In such applications, short data packets should be delivered with a minimum reliability of 99.999\% and an end-to-end (E2E) delay of less than 10-150\,ms \cite{salehi2023URLLC, peng2023resource}.    

In this paper, we study a cell-free massive MIMO system with URLLC user equipments (UEs) and multi-static sensing in a cluttered environment.
Both sensing and communication signals contribute to the sensing task. However, the sensing signals can cause interference for the UEs. Thus, we design the precoding vectors to null the interference only for the UEs. 
We consider a maximum DEP threshold, representing the reliability requirement, together with a maximum transmission delay threshold as the URLLC requirements. Moreover,
a minimum signal-to-interference-plus-noise ratio (SINR) is considered as the sensing requirement. This is motivated by the fact that given a fixed false alarm probability, detection probability increases with a higher sensing SINR \cite{behdad2023}, \cite[Chap. 3 and 15]{richards2010principles}. Our aim is to maximize the energy efficiency (EE) of downlink transmission. This involves minimizing power consumption for a given blocklength in line with URRLC requirements.
The main contributions of this paper are as follows:\begin{itemize}
\item We derive an upper bound for the DEP and transmission delay in finite blocklength cell-free massive MIMO communications. 
\item We propose a power allocation design approach, called \textit{SeURLLC+}, which maximizes EE while satisfying ISAC and URLLC requirements with dedicated sensing symbols coherently transmitted from ISAC APs. We cast the non-convex optimization problem in a form that facilitates the utilization of the Feasible Point Pursuit - Successive Convex Approximation (FPP-SCA) \cite{mehanna2014feasible}. 
\item We present a new definition of network availability accounting for both ISAC and URLLC requirements in the finite blocklength regime. We quantify this network availability by checking the feasibility of the proposed FPP-SCA-based algorithm. 
\item We investigate the impact of blocklength, sensing SINR threshold, and DEP threshold on network availability and EE by comparing the performance of the proposed power allocation algorithm with two benchmarks: i) \textit{URLLC-only}, representing a power allocation algorithm which aims to maximize EE by taking into account only URLLC requirements; ii) \textit{SeURLLC}, which represents the proposed power allocation algorithm without dedicated sensing symbols.
\end{itemize}

\section{System Model}
\label{section2}
We consider an ISAC cell-free massive MIMO system in URLLC scenarios. The system adopts a centralized radio access network (C-RAN) architecture \cite{cell-free-book} for uplink channel estimation and downlink communication, as well as multi-static sensing as in \cite{behdad2023} where more than two APs are involved in sensing. All the APs are connected through fronthaul links to the edge cloud and are fully synchronized.
 We consider the original form of cell-free massive MIMO \cite{ lancho2023cell}, where
all the $N_{\rm tx}$ ISAC APs jointly serve the $N_{\rm ue}$ URLLC UEs by transmitting precoded signals containing both communication and sensing symbols.  
Meanwhile, the $N_{\rm rx}$ sensing receive 
APs simultaneously sense the candidate location to detect if there is any object of interest. Each AP is equipped with an array of $M$ antennas arranged in a horizontal uniform linear array (ULA) with half-wavelength spacing.
The respective array response vector is $
       \textbf{a}(\varphi,\vartheta) =\begin{bmatrix}
          1\!&\! e^{j\pi \sin(\varphi)\cos(\vartheta)}& \ldots& e^{j(M-1)\pi\sin(\varphi)\cos(\vartheta)}
        \end{bmatrix}^T,
       $
 where $\varphi$ and $\vartheta$ are the azimuth and elevation angles from the AP to the target location, respectively \cite{bjornson2017massive}. 

We consider narrowband communications and the finite blocklength regime for URLLC UEs, where a packet of $b_i$ bits is sent to UE $i$ within a transmission block with blocklength $L=L_p+L_d$ symbols using the coherence bandwidth $B$. $L_p$ and $L_d$ are the number of symbols for pilot and data, respectively. It is expected that duration of each URLLC transmission, denoted by $T$, is shorter than one coherence time $T_ c$, i.e., $T<T_ c$ \cite{nasir2021cell}. 
However, without loss of generality, we assume $T=T_ c$ so that we estimate the channel for each transmission. 

The communication channels are modeled as spatially correlated Rician fading which are assumed to remain constant during each coherence block, and the channel realizations are independent of each other. Let $\textbf{h}_{i,k}\in \mathbb{C}^M$ denote the channel between ISAC AP $k$ and UE $i$, modeled as
\begin{align}
    \textbf{h}_{i,k} =e^{j\psi_{i,k}}\bar{\textbf{h}}_{i,k}\,+\,\tilde{\textbf{h}}_{i,k},\vspace{-2mm}
\end{align}
which consists of a semi-deterministic line-of-sight (LoS) path, represented by $e^{j\psi_{i,k}}\bar{\textbf{h}}_{i,k}$ with unknown phase-shift $\psi_{i,k}\sim \mathcal{U}[0,2\pi)$ (uniformly distributed on $[0,2\pi)$), and a stochastic non-line-of-sight (NLoS) component $\tilde{\textbf{h}}_{i,k}\sim \mathcal{CN}(\textbf{0},\tilde{\textbf{R}}_{i,k})$ with the spatial correlation matrix $\tilde{\textbf{R}}_{i,k}$. The $\bar{\textbf{h}}_{i,k}$ and $\tilde{\textbf{R}}_{i,k}$ include the combined effect of geometric path loss and shadowing.
We concatenate the channel vectors $\textbf{h}_{i,k}$ in the collective channel vector $\textbf{h}_{i}=\begin{bmatrix}
\textbf{h}_{i,1}^T& \ldots&
\textbf{h}_{i,N_{\rm tx}}^T
\end{bmatrix}^T\in \mathbb{C}^{N_{\rm tx}M}$ for UE $i$. 

Let $s_i$ and $s_0$ represent the downlink communication symbol for UE $i$ and sensing symbol, respectively. The symbols are independent and have zero mean and unit power. Moreover, let $\rho_i\geq 0$ and $\rho_0\geq 0$ be respectively the power control coefficients for UE $i$ and the target, which are fixed throughout the transmission.
Then, the transmitted signal from transmit AP $k\in\{1, \cdots, N_{\rm tx}\}$ at time instance $m$ is
\begin{equation}\label{x_k}
\vspace{-1mm}
    \textbf{x}_k[m]= \sum_{i=0}^{N_{\rm ue}} \sqrt{\rho_{i}}\textbf{w}_{i,k} s_{i}[m]=\textbf{W}_k \textbf{D}_{\rm s}[m]\boldsymbol{ \rho},
    \vspace{-1mm}
\end{equation}
where the vectors
 $\textbf{w}_{i,k}\in \mathbb{C}^{M}$ and $\textbf{w}_{0,k}\in \mathbb{C}^M$ are the transmit precoding vectors for transmit AP $k$ corresponding to UE $i$ and the sensing signal, respectively.
In \eqref{x_k}, $\textbf{W}_k= \begin{bmatrix}
\textbf{w}_{0,k} & \textbf{w}_{1,k} & \cdots & \textbf{w}_{N_{\rm ue},k}
\end{bmatrix}$, ${\textbf{D}}_{\rm s}[m]=\textrm{diag}\left(s_0[m],s_1[m],\ldots,s_{N_{\rm ue}}[m]\right)$ is the diagonal matrix containing the sensing and communication symbols, and $\boldsymbol{\rho}=[\sqrt{\rho_0} \ \ldots \sqrt{\rho_{N_{\rm ue}}}]^T$.

Let $n_i[m]\!\sim\!\mathcal{CN}(0,\sigma_n^2)$ denote the receiver noise at UE $i$ at time instance $m$ and the collective precoding vectors ${\textbf{w}_i\!=\![
        \textbf{w}_{i,1}^T\, \textbf{w}_{i,2}^T\, \hdots\, \textbf{w}_{i,N_{\rm tx}}^T
    ]^T\in \mathbb{C}^{N_{\rm tx}M}}$ and $\textbf{w}_0\!=\![\textbf{w}_{0,1}^T\,\textbf{w}_{0,2}^T\,\hdots \,\textbf{w}_{0,N_{tx}}^T]^T\!\in\!\mathbb{C}^{N_{\rm tx}M}$ be the centralized precoding vectors for UE $i$ and sensing location, respectively. Finally, the received signal at UE $i$ is given by
\begin{align}    \label{y_i-2}
     y_i[m] =&\underbrace{\sqrt{\rho_i}\textbf{h}_{i}^{H}\textbf{w}_{i} s_{i}[m]}_{\textrm{Desired signal}}+ \underbrace{\sum_{j=1,j\neq i}^{N_{\rm ue}}\sqrt{\rho_j}\textbf{h}_{i}^{H}\textbf{w}_{j} s_{j}[m]}_{\textrm{Interference signal due to the other UEs}}\nonumber\\
    &+ \underbrace{\sqrt{\rho_0}\textbf{h}_{i}^{H}\textbf{w}_{0} s_{0}[m]}_{\textrm{Interference signal due to the sensing}}+ \underbrace{n_i[m]}_{\textrm{Noise}}.
\end{align}

Using the independence of the data and sensing signals, the average transmit power for transmit AP $k$ is
\begin{equation} \label{eq:Pk}
    P_k = \sum_{i=0}^{ N_{\rm ue}}\rho_i\mathbb{E}\left\{\Vert \textbf{w}_{i,k} \Vert^2\right\}, \quad k=1,\ldots,N_{\rm tx} 
\end{equation}
 which should not exceed the maximum power limit $P_{\rm tx}$.

\subsection{Multi-Static Sensing}
We consider multi-static sensing where multiple distributed transmit and receive APs are involved. 
 The network senses the target location during downlink phase. We assume there is a LoS connection between the target location
and each transmit/receive AP. 
In the presence of the target, each receive AP receives the reflected signals from the target, as well as, undesired signals, known as clutter. The clutter is independent of the presence of the target and can be considered as interference for sensing. However, thanks to the C-RAN architecture, the LoS signals between transmit and receive APs are known and can be cancelled out at the edge cloud, since the transmitted signals and the AP locations are known. Therefore, the interference  signals correspond only to the reflected paths through the obstacles and are henceforth referred to as target-free channels.

Let $\textbf{H}_{r,k}\in \mathbb{C}^{M \times M}$ denote the target-free channel matrix between transmit AP $k$ and receive AP $r$. 
We use the correlated Rayleigh fading model for the NLoS channels $\textbf{H}_{r,k}$, which is modeled using the Kronecker model \cite{behdad2023}
. We define the random matrix $\textbf{W}_{{\rm ch},(r,k)}\in \mathbb{C}^{M \times M}$ with independent and identically distributed (i.i.d.) entries with $\mathcal{CN}(0,1)$ distribution. The matrix $\textbf{R}_{{\rm rx},(r,k)} \in \mathbb{C}^{M \times M}$ represents the spatial correlation matrix corresponding to receive AP $r$ and with respect to the direction of transmit AP $k$. Similarly, $\textbf{R}_{{\rm tx},(r,k)}\in \mathbb{C}^{M \times M}$ is the spatial correlation matrix corresponding to transmit AP $k$ and with respect to the direction of receive AP $r$. Then, the channel $\textbf{H}_{r,k}$ is 
\vspace{-2mm}
\begin{equation} \label{eq:Hrk}
    \textbf{H}_{r,k} = \textbf{R}^{\frac{1}{2}}_{{\rm rx},(r,k)} \textbf{W}_{{\rm ch},{(r,k)}}\left(\textbf{R}^{\frac{1}{2}}_{{\rm tx},(r,k)}\right)^T,
    \vspace{-1mm}
\end{equation}
 where the channel gain is determined by the geometric path loss and shadowing, and is included in the spatial correlation matrices.
In the presence of the target, the received signal at AP $r$ and time instance $m$, is 
\begin{align}\label{y_rPrim}
          &\textbf{y}_r[m]
           =\sum_{k=1}^{N_{\rm tx}}\underbrace{\alpha_{r,k}\sqrt{\beta_{r,k} }\textbf{a}(\phi_{0,r},\theta_{0,r})\textbf{a}^{T}(\varphi_{0,k},\vartheta_{0,k})\textbf{x}_k[m]}_{\textrm{desired reflections from the target}}\nonumber\\
          &+\sum_{k=1}^{N_{\rm tx}}\underbrace{\textbf{H}_{r,k}\textbf{x}_k[m]}_{\textrm{undesired signals}}+\textbf{n}_r[m], \quad m=1, \ldots, L_d
\end{align}
where $\textbf{n}_r[m]\sim \mathcal{CN}(\textbf{0},\sigma_n^2\textbf{I}_M)$ is the receiver noise at the $M$ antennas of receive AP $r$. 
Here,  $\beta_{r,k}$ is the channel gain including the path loss from transmit AP $k$ to receive AP $r$ through the target and the variance of bi-static radar cross-section (RCS) of the target denoted by $\sigma_{\rm rcs}$. $\alpha_{r,k}\!\sim\!\mathcal{CN}(0,\,1)$ is the normalized RCS of the target for the respective path. 
We assume the RCS values are i.i.d., and follow the Swerling-I model meaning that they are constant throughout the consecutive $L_d$ symbols collected for sensing.
Each receive AP sends their respective signals $\textbf{y}_r[m]$, for $r=1,\ldots,N_{\rm rx}$, to the edge cloud, to form the collective received signal
    $\label{y-prime}
  \textbf{y}[m]= \begin{bmatrix}
    \textbf{y}_{1}^T[m]&  \cdots&\textbf{y}_{N_{\rm rx}}^T[m]
    \end{bmatrix}^T 
  $.
  
According to  \cite{behdad2023}, the sensing SINR for the received vector signal over $L_d$ symbols 
is
\begin{align} \label{gamma_s_main}
     \mathsf{SINR}_{\rm s}= \frac{\boldsymbol{\rho}^T \textbf{A} \boldsymbol{\rho}}{L_d MN_{\rm rx}\sigma_n^2+ \boldsymbol{\rho}^T\textbf{B}\boldsymbol{\rho} },
\end{align}
where the positive semi-definite matrices $\textbf{A}$ and $\textbf{B}$ can be computed similarly to \cite[Eqs.~(39)-(40)]{behdad2023} with respect to i.i.d. RCS values as follows
\begin{align}
    \textbf{A}=& M\,\sum_{m=1}^{L_d}\textbf{D}_{\rm s}^H[m] \Bigg(\sum_{r=1}^{N_{\rm rx}}\sum_{k=1}^{N_{\rm tx}}\beta_{r,k}\textbf{W}_k^H\textbf{a}^{*}_k\textbf{a}^{T}_k \textbf{W}_k\Bigg)\textbf{D}_{\rm s}[m]\\
\textbf{B}=&\sum_{m=1}^{L_d}\textbf{D}_{\rm s}^H[m] \left(\sum_{r=1}^{N_{\rm rx}}\sum_{k=1}^{N_{\rm tx}}\mathrm{tr}\left(\textbf{R}_{\mathrm{rx},(r,k)}\right)\textbf{W}_k^H \textbf{R}^{T}_{{\rm tx},(r,k)}\textbf{W}_k \right)\nonumber\\
&\hspace{3mm}\times\textbf{D}_{\rm s}[m],
\end{align}
where $\textbf{a}_k=\textbf{a}(\varphi_{k},\vartheta_{k})$.

\subsection{Transmit ISAC Precoding Vectors}\label{section:precder}
 The communication and sensing transmit precoding vectors are obtained based on regularized zero forcing (RZF) and zero forcing (ZF) approaches, respectively. The unit-norm RZF precoding vector for UE $i$ is given as $\textbf{w}_{i}=\frac{\bar{\textbf{w}}_{i}}{\left \Vert \bar{\textbf{w}}_{i}\right \Vert}$, with \begin{align}
     \bar{\textbf{w}}_{i}
     \!=\!\left(\sum\limits_{j=1}^{N_{\rm ue}}\hat{\textbf{h}}_j\hat{\textbf{h}}_j^H+\delta\textbf{I}_{N_{\rm tx}M}\right)^{\!\!-1}\!\!\hat{\textbf{h}}_i,\quad i=1,\ldots,N_{\rm ue},  
\end{align}
where $\delta$ is the regularization parameter, and $\hat{\textbf{h}}_{j}=\begin{bmatrix}
\hat{\textbf{h}}_{j,1}^T& \ldots&
\hat{\textbf{h}}_{j,N_{\rm tx}}^T
\end{bmatrix}^T\in \mathbb{C}^{N_{\rm tx}M}$ 
is the linear minimum mean-squared error channel estimate of the communication channel $\textbf{h}_{j}$, obtained as in \cite{wang2020uplink}. 
If the number of UEs is larger than the number of mutually orthogonal pilot sequences, then each pilot sequence may be assigned to multiple UEs using the pilot assignment algorithm in \cite[Algorithm 4.1]{cell-free-book}.

We aim to null the destructive interference from the sensing signal to the UEs by using the unit-norm ZF sensing precoding vector $\textbf{w}_{0} = \frac{\bar{\textbf{w}}_{0}}{\left \Vert \bar{\textbf{w}}_{0}\right \Vert}$, where 
   $\bar{\textbf{w}}_{0}=\left(\textbf{I}_{N_{\rm tx}M}-\textbf{U}\textbf{U}^{H}\right)\textbf{h}_{0}$,
and $\textbf{U}$ is the unitary matrix with the orthogonal columns that span the column space of the matrix $\begin{bmatrix}
         \hat{\textbf{h}}_{1}&  \ldots& \hat{\textbf{h}}_{N_{\rm ue}}
        \end{bmatrix}$. $\textbf{h}_{0}= \begin{bmatrix}
\sqrt{\beta_{0,1}}\textbf{a}^T(\varphi_{0,1},\vartheta_{0,1})&  \ldots&
\sqrt{\beta_{0,N_{\rm tx}}}\textbf{a}^T(\varphi_{0,N_{\rm tx}},\vartheta_{0,N_{\rm tx}})
\end{bmatrix}^T\in \mathbb{C}^{N_{\rm tx}M}$ is the concatenated sensing channel between all the ISAC transmit APs and the target, including the corresponding path loss and the array response vectors $\textbf{a}(\varphi_{0,k},\vartheta_{0,k})$.

\section{Reliability and Delay Analysis for URLLC}\label{section3}
In this section, we derive an upper bound on the DEP and the transmission delay which are considered as the URLLC requirements. In the finite blocklength regime, the communication data cannot be transmitted without error.
From \cite{ren2020joint}, ergodic data rate of UE $i$ can be approximated as
\begin{equation}\label{eq:R_i}
     R_i \! \approx \!
     \mathbb{E}\Bigg\{\!\!\left(1\!-\!\beta\right)\! \log_2{\left(1+ \mathsf{SINR}^{\rm (dl)}_{i}\right)}\!-\!\frac{Q^{-1}(\epsilon_{i})}{\ln(2)}\!\sqrt{\!\frac{(1\!-\!\beta)V_i}{L}}\Bigg\},
     \vspace{-0mm}
\end{equation}
where $\beta = \frac{L_p}{L}$, $\epsilon_i$ denotes the DEP when transmitting $b_i$ bits to UE $i$, $\mathsf{SINR}^{\rm (dl)}_{i}$ is the instantaneous downlink communication SINR for UE $i$,  $V_i= 1-\left(1+ \mathsf{SINR}^{\rm (dl)}_{i}\right)^{-2}$ is the channel dispersion, and $Q(\cdot)$ refers to the Gaussian Q-function. Due to the fact that $V_i<1$, the ergodic data rate can be lower bounded by 
\begin{align}\label{eq:R_i_lb}
     R_i & \!\geq \!   \left(1\!-\!\beta\right)\!\mathbb{E}\Bigg\{\!\! \log_2{\left(1+ \mathsf{SINR}^{(\rm dl)}_i\right)}\!\!\Bigg\}\!-\!\frac{Q^{-1}(\epsilon_{i})}{\ln(2)}\sqrt{\frac{(1-\beta)}{L}}.%
\end{align}
Moreover, given that only $\mathbb{E}\left\{\textbf{h}_{i}^H \textbf{w}_{i}\right\}$ is known at UE $i$, and according to \cite[Thm.~6.1]{cell-free-book} and \cite[Lem.~1]{behdad2023}, 
\begin{align}\label{E_sinr}
   \mathbb{E} \left\{\log_2{\left(1+ \mathsf{SINR}^{\rm (dl)}_{i}\right)}\right\} \geq \log_2{\left(1+ \overline{\mathsf{SINR}}^{(\rm dl)}_i\right)}
\end{align}
where 
\begin{align}\label{sinr_i}
&\overline{\mathsf{SINR}}^{(\rm dl)}_i=
    \frac{\rho_i\mathsf{b}_i^2}{\sum_{j=0}^{{\rm N_{ue}}}\rho_j\mathsf{a}_{i,j}^2+\sigma_n^2}, \quad i= 1,\ldots,N_{\rm ue}
    \end{align}
    with
   \vspace{-2mm}
\begin{align}
    & \mathsf{b}_i = \left \vert \mathbb{E}\left\{\textbf{h}_{i}^H \textbf{w}_{i}\right\}\right\vert,\quad \mathsf{a}_{i,i} =\sqrt{\mathbb{E}\left\{\left \vert\textbf{h}_{i}^H \textbf{w}_{i}\right\vert^2\right\}- \mathsf{b}_i^2} \\
    &
    \mathsf{a}_{i,j} = \sqrt{\mathbb{E}\left\{\left \vert\textbf{h}_{i}^H \textbf{w}_{j}\right\vert^2\right\}},\quad j= 0,1,\ldots,N_{\rm ue}, \quad j\neq i.
    \end{align}
The expectations are taken with respect to the random channel realizations. Now, using \eqref{E_sinr}
and substituting $R_i = \frac{b_i}{L}$ into \eqref{eq:R_i_lb}, we obtain an upper bound for the DEP as
\begin{align}\label{upperbound}
    \epsilon_{i}\leq\epsilon_{i}^{\rm (ub)}\! \triangleq\!Q\!\left(\sqrt{L-L_p} \left[\ln\left(1+ \overline{\mathsf{SINR}}^{(\rm dl)}_i\right)-\frac{b_i \ln{2}}{L-L_p} \right]\right).
\end{align}

In this paper, we focus on the transmission delay and leave the analysis of E2E delay as future work. Let $D_i^t$ denote the transmission delay of UE $i$, expressed as
\begin{align}\label{eq:delay}
    D_i^{\rm t}= \frac{T}{1-\epsilon_i}=\frac{L}{B(1-\epsilon_{i})}, \quad  i= 1,\ldots,N_{\rm ue}
\end{align}
 where $T=\frac{L}{B}$ is the time duration of one URLLC transmission with blocklength $L$ and $\frac{1}{1-\epsilon_i}$ is the average number of transmissions given that re-transmission is allowed. To satisfy the reliability requirement, $\epsilon_i$ should be less than the maximum tolerable DEP denoted by $\epsilon_i^{(\rm th)}$. Then, since $\epsilon_i \leq\epsilon_{i}^{\rm (ub)}\leq \epsilon_{i}^{\rm (th)} $, the transmission delay is upper-bounded as
\begin{equation}\label{eq:D_th}
    D_i^{\rm t}\leq \frac{L}{B\left(1-\epsilon_{i}^{\rm (th)}\right)}\triangleq D_i^{\rm (ub)}\leq  D_i^{\rm (th)},
\end{equation}
where $D_i^{\rm (th)}$ is the maximum tolerable delay by UE $i$ and $D_i^{\rm (ub)}\leq  D_i^{\rm (th)}$ should be satisfied to guarantee the delay requirement. This implies that the blocklength cannot exceed $D^{\rm (th)}_i B (1-\epsilon_i^{\rm (th)})$. Thus, we can define the maximum tolerable blocklength by $L_{\rm max}$, where
\begin{equation}\label{L_max}
    L_{\rm max} = \min \left\{ D^{\rm (th)}_i B \left(1-\epsilon_i^{\rm (th)}\right)|  i=1, \cdots, N_{\rm ue}\right\}.
\end{equation}

\section{Power Allocation and Network Availability}\label{section4}

\subsection{Power Allocation Algorithm}
  We propose a power allocation algorithm that aims to maximize EE while satisfying the per-AP power constraints along with the URLLC and sensing requirements. 
  We have numerically shown in \cite{behdad2022power} and \cite{behdad2023}, that maximizing the sensing SINR improves the target detection probability under a constant false alarm probability. When studying other sensing tasks, it is naturally desired to keep the sensing SINR above a required threshold. Motivated by this fact, the optimization problem can be cast as
\begin{subequations} 
\begin{align} 
    \underset{\boldsymbol{\rho}\geq \textbf{0}}{\textrm{maximize}} \quad & \mathsf{EE} = \frac{B\,\sum_{i=1}^{N_{ue}} b_i}{L_d\,P_{\rm total}/(1-\epsilon_{\rm th})}, \quad \textrm{bits/Joule}\label{obj_func}\\
   \textrm{subject to} \quad & \epsilon_i \leq \epsilon_{i}^{\rm (ub)}\leq\epsilon_{i}^{\rm (th)},\quad i=1,\cdots,N_{\rm ue}\label{max:cona}\\
   &\mathsf{SINR}_{\rm s}\geq \gamma_{\rm s},\label{max:conc}\\
    & P_k \leq P_{\rm tx},\quad\quad \quad \quad k=1,\cdots,N_{\rm tx} \label{max:cond}
    \end{align}
\end{subequations}
where $\epsilon_{\rm th}=\max\{\epsilon_i^{\rm th}| \forall i\}$ is the maximum DEP threshold. In fact, we use $\epsilon_{\rm th}$ to compute the average transmission time in \eqref{eq:delay}, instead of using the individual DEP for each UE. This gives us a lower bound on the average EE as in \eqref{obj_func} where $P_{\rm total}= \sum_{k=1}^{N_{\rm tx}}P_{k}$ is the average total power consumption. Due to the unit-power centralized precoding vectors, $P_{\rm total}=\sum_{j=0}^{N_{\rm ue}}\rho_j= \Vert \boldsymbol{\rho}\Vert^2$. The $\gamma_{\rm s}$ is the required sensing SINR that should be selected according to the sensing task and other parameters such as $L_d$ in this work that represents the sensing duration. $P_{\rm tx}$ is the maximum transmit power for each AP.

Given that the total number of information bits, i.e., $\sum_{i=1}^{N_{ue}} b_i$, the bandwidth $B$, the blocklength $L$, and the maximum DEP $\epsilon_{\rm th}$ are constant, maximizing the lower bound of $\mathsf{EE}$ is equivalent to minimizing the total power consumption from all ISAC transmit APs. 

According to \eqref{upperbound}, the reliability constraints in \eqref{max:cona} can be written as communication SINR constraints in the form
\begin{align}\label{cons_SE2}
\overline{\mathsf{SINR}}^{\rm (dl)}_{i}\! \geq \gamma_{{\rm c},i}\triangleq\exp\left(
 \frac{Q^{-1}(\epsilon_i^{(\rm th)})}{\sqrt{L-L_p}}+\frac{b_i \ln{2}}{L-L_p}\right)-1
\end{align}
where $\gamma_{{\rm c},i}$ is the SINR threshold for UE $i$ which is a function of the DEP threshold, blocklength, number of pilot symbols, and packet size. From \eqref{sinr_i}, the communication SINR constraints 
can be rewritten as second-order cone (SOC) constraints in terms of $\boldsymbol{\rho}=[\sqrt{\rho_0}\ \ldots \sqrt{\rho_{N_{\rm ue}}}]^T$  as
\begin{align}
    &\left \Vert \begin{bmatrix}  \mathsf{a}_{i,0}\sqrt{\rho_{0}} & \mathsf{a}_{i,1}\sqrt{\rho_{1}} & \ldots & \mathsf{a}_{i,N_{\rm ue}}\sqrt{\rho_{N_{\rm ue}}} &
    \sigma_n
    \end{bmatrix}   \right \Vert \nonumber\\
    &\leq
    \frac{\sqrt{\rho_i} \mathsf{b}_{i}}{\sqrt{\gamma_{{\rm c},i}}}, \quad i= 1,\ldots,N_{\rm ue}.  \label{conb2}
\end{align}

The per-AP power constraints in \eqref{max:cond} can also be rewritten in SOC form  as
\begin{align}
\left \Vert \textbf{F}_k\boldsymbol{\rho}\right \Vert \leq \sqrt{P_{\rm tx}}, \quad k=1,\ldots,N_{\rm tx}, \label{conc2}
\end{align}
where $\textbf{F}_k =\textrm{diag}\left( \sqrt{\mathbb{E}\left\{\Vert\textbf{w}_{0,k}[m]\Vert^2\right\}}, \ldots,\sqrt{\mathbb{E}\left\{\Vert \textbf{w}_{N_{\rm ue},k}\Vert^2\right\}}\right)$. Using \eqref{gamma_s_main}, the sensing constraints in \eqref{max:conc} are expressed as
\begin{equation}\label{conb3}
    \boldsymbol{\rho}^T\!\left(\gamma_{\rm s} \textbf{B}-\textbf{A}\right)\boldsymbol{\rho}\!\leq \!-\gamma_{\rm s}L_d MN_{\rm rx}\sigma_n^2,
\end{equation}
which is not convex. 
To handle this non-convexity, we use FPP-SCA method \cite{mehanna2014feasible}. To avoid any potential infeasibility issue during the initial iterations of the algorithm, we add slack
variable $\chi\geq0$ and a slack penalty to the convexified problem at each iteration. However, to have the fully satisfied constraints, the slack variable should be zero at the end. Thus, we set the slack variable to zero in the next iterations if they are lower than a threshold denoted by $\epsilon_{\chi}>0$.
Thus, the sensing constraints at the $(c+1)^{\rm th}$ iteration of the FPP-SCA algorithm are
\begin{align}
    &\gamma_{\rm s}\boldsymbol{\rho}^T \textbf{B}\boldsymbol{\rho}+ 2\Re\left(\left(\boldsymbol{\rho}^{(c)}\right)^T \left(-\textbf{A}\right)\boldsymbol{\rho}\right)\nonumber\\
    & \leq -\gamma_{\rm s}L_d\,MN_{\rm rx}\sigma_n^2+\left(\boldsymbol{\rho}^{(c)}\right)^T \left(-\textbf{A}\right)\boldsymbol{\rho}^{(c)} + \chi,  \label{cond2}
\end{align}
 where $\boldsymbol{\rho}^{(c)}$ is the optimal solution found at the $c^{\rm th}$ iteration. Then, the convex problem that is solved at the  $(c+1)^{\rm th}$ iteration becomes 
\begin{align} \label{EE optimization}
    \underset{\boldsymbol{\rho}\geq \textbf{0}, \chi\geq 0}{\textrm{minimize}} \quad & f= \Vert \boldsymbol{\rho}\Vert+\lambda \chi\\
   \textrm{subject to} \quad  & \eqref{conb2}, \eqref{conc2}, \eqref{cond2}\nonumber
  \end{align}
where $f$ is the objective function and $\lambda>0$ is the penalty coefficient. The steps of FPP-SCA procedure are outlined in Algorithm~\ref{alg:FPP-SCA}.
\vspace{-2mm}
\begin{algorithm}
	\caption{FPP-SCA Procedure for Power Allocation Problem in \eqref{EE optimization}.} \label{alg:FPP-SCA}
	   \begin{algorithmic}[1]
		\State {\bf Initialization:} Set an arbitrary $\boldsymbol{\rho}^{(0)}\geq \textbf{0}$ and the solution accuracy $\epsilon>0$, $\epsilon_{\chi}>0$, and $\lambda>0$. Set $f^{(0)}\!= \!\infty$ and the iteration counter to $c=0$, and a maximum iteration $c_{\rm max}$.
        \State Set $\boldsymbol{\rho}^{(c+1)}$ to the solution of the convex problem in \eqref{EE optimization},  where the previous iterate $\boldsymbol{ \rho}^{(c)}$ is taken as constant.
        \State If $\chi <\epsilon_{\chi}$, set $\chi = 0$ for the next iteration.
        \State $c\leftarrow c+1$.
        \State If $|f^{(c)}-f^{(c-1)}|> \epsilon$ and $c\leq c_{\rm max}$, go to Step 2, otherwise go to Step 6.
        \State If $\chi \neq 0$, mark the problem as infeasible. 
	\State {\bf Output:} The transmit power coefficients  $\boldsymbol{\rho}^{(c)}$.
	    \end{algorithmic}
 \end{algorithm}
 \vspace{-4mm}
\subsection{Network Availability}
Network availability is a performance metric that measures the probability of providing the required quality of service (QoS) for UEs. 
However, this metric is relative and can vary for ISAC systems. For example, in sensing-based mission-critical applications, sensing performance plays a crucial role in serving URLLC UEs. Hence, we define the network availability as the probability that reliability and delay requirements for URLLC, as well as a minimum SINR for sensing are satisfied. 

We calculate network availability using Monte Carlo simulations to compute the probability that the transmit power coefficients, obtained from our proposed power allocation algorithm, satisfy all constraints including URLLC, sensing, and per-AP power constraints, i.e., the problem is feasible.

\section{Numerical Results}\label{section5}
In this section, we present numerical results to evaluate the performance of the proposed power allocation algorithm.  The $N_{\rm tx}=16$ ISAC transmit APs are uniformly distributed in an area of 500\,$\rm m \times$\,500\,$\rm m$. The sensing location is in the center of the area, i.e., $(250,250)$.
The number of antenna elements per AP is $M=4$.
The $N_{\rm rx}=2$ receive APs are located close to the sensing location at $(200,250)$ and  $(300,250)$. We consider a total of $N_{\rm ue}=8$ URLLC UEs in the network. We set $P_{\rm tx}=100$\,mW and the uplink pilot transmission power is $50$\,mW.

The large-scale fading coefficients, shadowing
parameters, probability of LoS, and the Rician factors are
simulated based on the 3GPP Urban Microcell model, defined in \cite[Table B.1.2.1-1, Table B.1.2.1-2, Table B.1.2.2.1-4]{3gpp2010further}. The path loss for the Rayleigh fading target-free channels is also modeled by the 3GPP Urban Microcell model with the difference that the channel gains are multiplied by an additional scaling parameter equal to $0.3$ to suppress the known parts of the target-free channels due to the permanent obstacles \cite{behdad2023}. The sensing channel gains are computed by the bi-static radar range equation \cite{richards2010principles} with $\sigma_{\rm rcs} = 0\,\rm dBsm$. The RCS of the target is modeled by the Swerling-I model. The carrier frequency, the bandwidth, and the noise variance are set to $1.9$\,GHz, $200$\,KHz, and $-114$\,dBm, respectively. The number of pilot symbols is $L_p = 10$.
The spatial correlation matrices for the communication channels are generated by using the local scattering model in \cite[Sec.~2.5.3]{cell-free-book}. For all the UEs, we set the same packet size, the maximum transmission delay, and the DEP threshold to $b_i=256$\,bits, $D_i^{\rm (th)}=1$\,ms, and $\epsilon_{\rm th}=10^{-5}$, respectively. In the proposed algorithm, we set $\epsilon=10^{-3}$, $\epsilon_{\chi}=10^{-6}$, $\lambda = 10$, and $c_{\rm max}=50$.

We compare the performance of the proposed power allocation algorithm, i.e., \textit{SeURLLC+}, with two benchmarks: i) \textit{URLLC-only}, corresponding to a power allocation algorithm which maximizes EE for URLLC without considering sensing requirements, and ii) \textit{SeURLLC}, which represents the proposed power allocation algorithm without additional sensing symbols.

\begin{figure}
\centering
\includegraphics[trim={1mm 0mm 13mm 8mm},clip,width=0.8\linewidth]{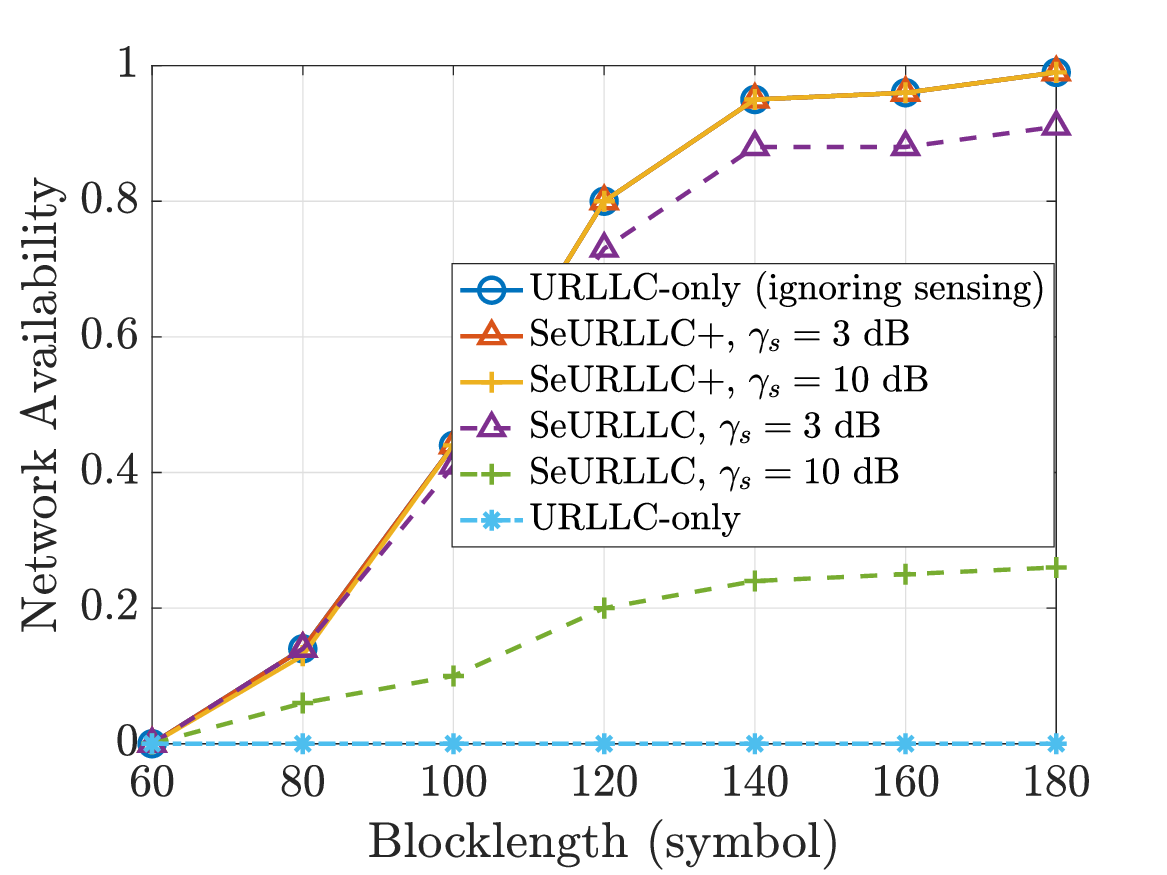}\vspace{-1.5mm}
   \caption{Network availability vs. blocklength.}
    \label{NA_BL}
      \vspace{-3mm}
  \end{figure}%
Fig.~\ref{NA_BL} shows the network availability versus the blocklength for two sensing SINR thresholds: $3$\,dB and\,$10$\,dB. Generally, the network availability increases with larger blocklengths but decreases when sensing is integrated, particularly when only communication symbols are employed. Notably, the \textit{URLLC-only} algorithm could not provide any sensing capability, highlighting the need for new power allocation algorithms to enable sensing in such systems. However, for both sensing requirements ($3$\,dB and $10$\,dB), the network availability with the proposed algorithm can approach the one of \textit{URLLC-only} with ignoring sensing in the system, only when additional sensing symbols are employed. The \textit{SeURLLC+} achieves network availability over 0.9 when the blocklength is  at least $140$\, symbols and reaches 0.99 when $L=180$. 
\begin{figure}
  \centering
  \includegraphics[trim={1mm 0mm 13mm 6mm},clip,width=0.8\linewidth]{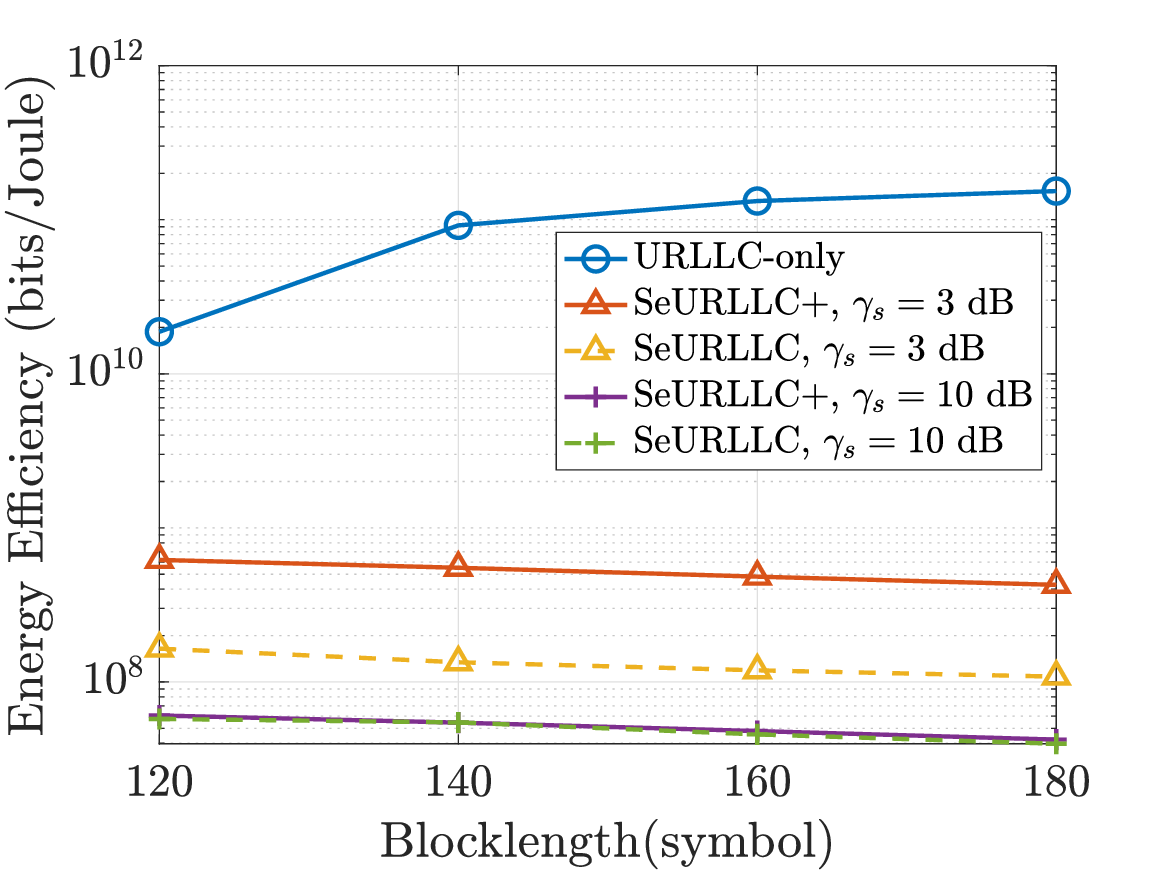}\vspace{-1.5mm}
    \caption{Energy efficiency vs. blocklength.}
    \label{EE_BL}\vspace{-3mm}
\end{figure}

Fig.~\ref{EE_BL} demonstrates the impact of blocklength on EE for two sensing SINR thresholds ($3$\,dB and $10$\,dB). The \textit{URLLC-only} algorithm is most energy-efficient, but it could not meet the sensing requirement (as in Fig.~\ref{NA_BL}). Increasing blocklength enhances the EE of \textit{URLLC-only}. In contrast, the EE for the proposed algorithm, both with and without sensing symbols, experiences a slight decrease with increasing blocklength. This phenomenon occurs due to the interplay between blocklength and total power, as power consumption is inherently linked to blocklength. 
Moreover, smaller sensing SINR thresholds lead to higher EE, as less power is needed for sensing. However, \textit{SeURLLC+} outperforms \textit{SeURLLC} by using dedicated sensing symbols which can result in up to around 75\% smaller power consumption due to smaller power demand on communication symbols to meet sensing criteria.  
\begin{figure}
\centering
\includegraphics[trim={1mm 0mm 12mm 6mm},clip,width=0.8\linewidth]{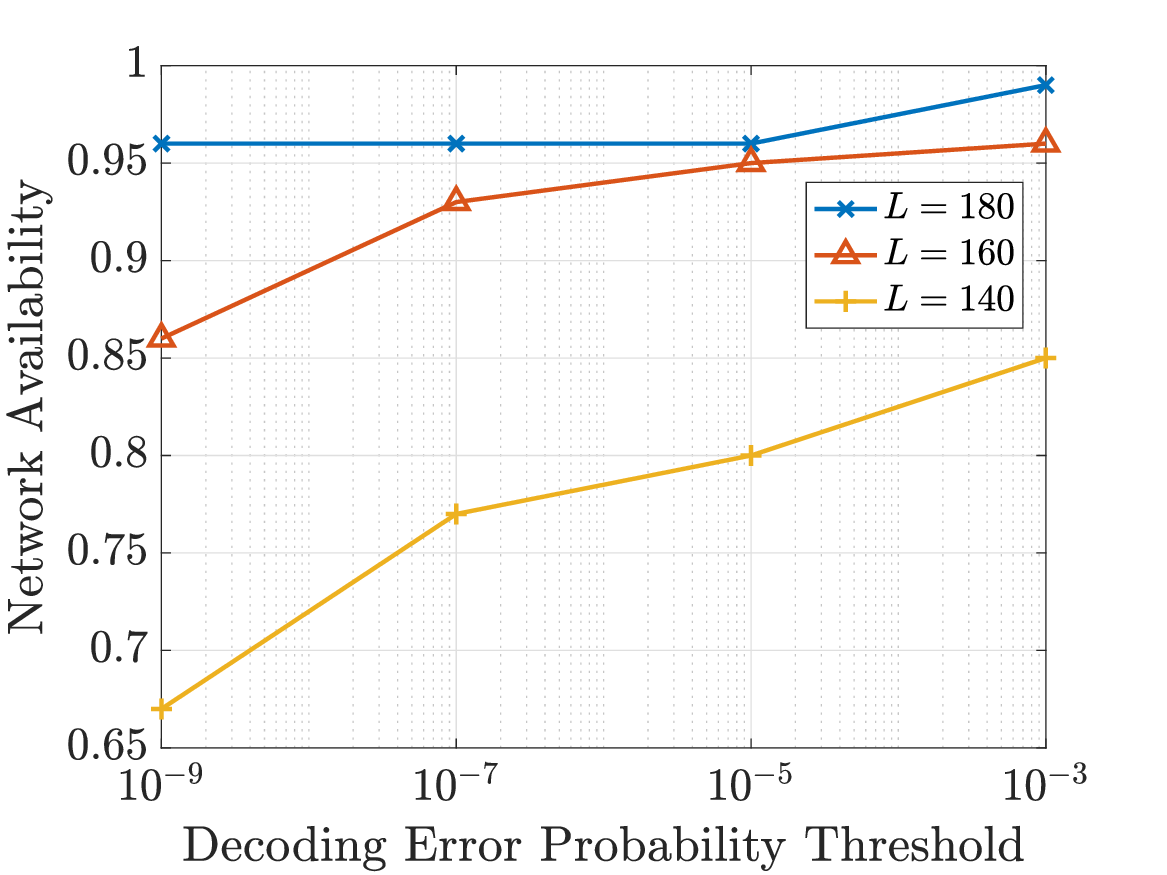}\vspace{-1.5mm}
   \caption{Network availability vs. decoding error probability (DEP) threshold for \textit{SeURLLC+}.}
    \label{NA_DEP}
      \vspace{-3mm}
\end{figure}

The impact of the reliability requirement, DEP threshold, on the network availability is shown in Fig.~\ref{NA_DEP}. 
We compare the results with three values of blocklength equal to 140, 160, and 180 symbols. The network availability increases as the DEP threshold grows and it is more sensitive as the blocklength decreases. Larger blocklength has higher network availability and is more robust against lower DEP thresholds, as we see the curves corresponding to the blocklength of 180 symbols are almost flat. 

\vspace{-2mm}

\section{Conclusion} \label{section6}
\vspace{-1mm}
In this paper, we proposed a power allocation algorithm for a downlink cell-free massive MIMO system with multi-static sensing and URLLC users. The algorithm maximizes EE while meeting the sensing and URLLC requirements. 
We introduced a new definition of network availability, which considers both ISAC and URLLC requirements and showed that power allocation algorithms without taking into account sensing requirements could not provide sensing. Moreover, the numerical results validated the proposed power allocation algorithm and showed that additional sensing symbols are needed to enhance the performance of the network. We observed that higher blocklength is more robust in terms of the network availability to satisfy stringent reliability requirements, although it would result in a lower energy efficiency.
\bibliographystyle{IEEEtran}
\bibliography{IEEEabrv.bib,refs.bib}
\end{document}